\documentclass[aps,pra,twocolumn,showpacs]{revtex4}
\usepackage{amsbsy,latexsym}
\usepackage{amsfonts}
\usepackage{amssymb}
\usepackage[mathscr]{eucal}

\newcommand{\carac}[1]{\chi_{#1}}
\newcommand{\al}{\alpha}

\newcommand{\de}{\delta}
\newcommand{\la}{\lambda}
\newcommand{\id}{\mathbb I}
\newcommand{\CF}{\mathsf n}

\newcommand{\tr}{{\rm tr}\,}

\newcommand{\ket}[1]{\vert #1 \rangle}
\newcommand{\bra}[1]{\langle #1 \vert}

\newcommand{\D}[3]{\mathfrak D^{(#1)}_{#2 #3}}
\newcommand{\Dc}[3]{\mathfrak D^{(#1) *}_{#2 #3}}

\begin{document}
\title{Entanglement assisted alignment of reference frames
using a dense covariant coding}
\author{E.~Bagan, M.~Baig, and R.~Mu{\~n}oz-Tapia}
\affiliation{Grup de F{\'\i}sica Te{\`o}rica \& IFAE, Facultat de
Ci{\`e}ncies, Edifici Cn, Universitat Aut{\`o}noma de Barcelona,
08193 Bellaterra (Barcelona) Spain}

\begin{abstract}
We present a procedure inspired by dense coding, which enables a
highly efficient transmission of information of a continuous
nature. The procedure requires the sender and the recipient to
share a maximally entangled state. We deal with the concrete
problem of aligning reference frames or trihedra by means of a
quantum system. We find the optimal covariant measurement and
compute the corresponding average error, which has a remarkably
simple close form. The connection of this procedure with that of
estimating unitary transformations on qubits is briefly discussed.
\end{abstract}

\maketitle

Entanglement has long been recognized as a powerful resource in
quantum communication. Teleportation~\cite{teleportation} and
dense coding~\cite{dense}, for instance, would not be possible
without entanglement. Even when entanglement is not strictly
necessary, one frequently runs across situations for which the use
of entangled states, instead of plain product states, provides a
significant improvement. Examples of this can be easily found in
the literature. This letter provides yet another interesting
instance, which one could refer to as dense covariant coding.

Two interesting problems in quantum communication 
in which entanglement plays a fundamental role are those of
sending 
the information that specifies \textit{(i)} a direction in space,
i.e., a unit vector $\vec n_1$,  or \textit{(ii)} three orthogonal
ones (a trihedron) $\CF=\{\vec n_{1}, \vec n_{2},\vec n_{3}\}$.
Whereas \textit{(i)} has been extensively discussed in the
literature~\cite{direction-1,direction-bbm,direction-2}, only
recently significant
attention~\cite{reference-ps,reference-bbm,ps2} has been paid
to~\textit{(ii)}. It has been shown that quantum states can indeed
be used to establish a common reference frame between two parties
(Alice and Bob). Thus, for instance, atoms or a number of spins
(throughout this letter we use the word spin as synonym of spin
1/2 particle) can encode the relative orientation of two trihedra.
The fidelity (or alternatively, the mean square error per axis) of
the optimal covariant communication protocol (where covariance
refers to the set of signal states being the orbit of a group;
$\rm SU(2)$ for the problem at hand) is now known for both, finite
and asymptotically large number $N$ of copies of the messenger
state.


In this letter we show that the intensive use of entanglement
yields a remarkable improvement over the approaches for aligning
spatial frames discussed above. More specifically, suppose Alice
and Bob share a maximally entangled state. Then, we will show that
using a covariant protocol it is possible to establish a common
reference frame with a mean square error per axis given by
$[1-\cos2\pi/(N+3)]/3$, which behaves as $2\pi^2/(3N^2)
$. This protocol bears a great similarity with dense coding as far
as the use it makes of entanglement and the remarkable improvement
it provides in the transmission of information~\cite{ariano}.
Dense coding has mainly been discussed for discrete signals.
However, the information we are attempting to transmit has an
intrinsically continuous nature: it refers to the relative
orientation of Alice and Bob and, in some
situations~\cite{direction-1,direction-bbm,direction-2,
reference-bbm,reference-ps,others, experiments, bruss}, such
information cannot be codified by a series of bits. Indeed,
 a digital representation of an orientation has
no meaning unless it is referred to a common reference frame. No
such frame will be assumed to be known to both  Alice and Bob
unless otherwise stated, though we will use Bob's to simplify the
mathematics. Hence, the messenger will have to be a quantum system
with intrinsic orientation. More specifically, in this letter we
will consider a system of spins. (See \cite{bartlett} for another
protocol of sending information without a shared reference frame.)
The subject of this letter is also related to the important issue
of estimating a unitary operation on qubits~\cite{ajv}. We will
come back to this point in the conclusions.

Suppose Alice and Bob have each of them a system of $N$ spins and
let us call ${\mathscr H}_A$ and ${\mathscr H}_B$ their respective
Hilbert spaces (throughout this letter subscripts {\em A} and {\em
B} will always refer to Alice and Bob).  Before they start their
intergalactic journeys, they prepare the state 
\begin{equation}
\ket{\Phi}=\sum_j a_j\ket{\Phi^j}=\sum_{j}{ a_j\over\sqrt{d_j}}
\sum_{m=-j}^j\ket{jm}_A\ket{jm}_B , \label{state Phi}
\end{equation}
where $j$ runs from zero to $N/2$ for $N$ even ($1/2$ to $N/2$ for
$N$ odd), $d_j=2j+1$ is the dimension of the representation~$\bf
j$ of $\rm SU(2)$, and $\sum_{j} a_{j}^2=1$. Also before
departure, they lock the orientation of their systems of $N$ spins
to that of their respective spacecrafts. When they are far apart,
they need to get aligned. Unfortunately,  their classical
computers crashed and they cannot retrieve the information about
the change of their relative orientation. At this point in time,
the state of Alice's and Bob's spins is still given by~(\ref{state
Phi}) but $\ket{jm}_A$ and  $\ket{jm}_B$ are now referred to
Alice's and Bob's reference frames respectively (in this
presentation the words spacecraft and reference frame are
synonyms). Relative to Bob's reference frame this state can be
written as
\begin{equation}
\ket{\Phi(g)}\equiv U_A(g)\otimes \id_B \ket{\Phi}, \label{Phi(g)}
\end{equation}
where $U_A(g)$ belongs to the direct sum of irreducible
representations of $\rm SU(2)$
and $g$ stands for the three Euler angles of the spacial rotation
that takes Bob's reference frame into Alice's. With no other
resource available, Alice sends her $N$ spins to~Bob, with the
hope that he will retrieve from them the information they need. To
do so, he is allowed to perform generalized collective
measurements on both Alice's and his own spins, namely, on the
state~(\ref{Phi(g)}).
Note that $\ket{\Phi}$ and $\ket{\Phi(g)}$ are maximally entangled
in each~$\bf j$. Note also that in~(\ref{state Phi}) all of these
representations appear only once, despite of the fact that in the
Clebsch-Gordan decomposition of $({\bf 1/2})^{\otimes N}$ they may
show up several times. We will show  that $\ket{\Phi}$ is optimal
for the problem at hand provided a suitable choice of~$a_j>0$ is
made (see Eq.~\ref{a_j} below).

%
The quality of the communication strategy can be quantified by the
averaged Holevo's error~\cite{holevo}
\begin{equation}
  \langle h\rangle=\sum_r \int dg\; h(g,g_r) p(r|g)    \label{error},
\end{equation}
where $ h(g,g')=\sum_{a=1}^3|\vec n_{a}(g)-\vec n_{a}(g')|^2$;
$\CF(g)=\{\vec n_{1}(g),\vec n_{2}(g),\vec n_{3}(g)\}$ defines the
frame Alice is transmitting to Bob; $\CF(g_r)=\{\vec
n_{1}(g_r),\vec n_{2}(g_r),\vec n_{3}(g_r)\}$ defines the frame
Bob guesses from the outcome $r$ of his measurement; and $dg$ is
the invariant Haar measure of $\mathrm{SU(2)}$. Each one of these
trihedra are labelled with the
%
parameters $g$  of the rotation that brings $\CF_0=\{\vec x,\vec
y,\vec z\}$ (the unit vectors along Bob's axes) into the desired
orientation. $p(r|g)$ is the conditional probability of Bob
obtaining the outcome $r$ if Alice sends $\CF(g)$.
Note that $h(g,g')$ is related to the character $\carac1$ of the
 representation $\bf1$ of $\rm SU(2)$ through
$h(g,g')=6-2\carac1(gg'^{-1})$. Hence, we just need to compute
$\langle\carac1\rangle$. {}From this, the square error per axis,
to which we referred above, is $(3-\langle\carac1\rangle)/6$.
Quantum Mechanics tells us that the conditional probability is
$p(r|g)=\bra{\Phi(g)}  O_r \ket{\Phi(g)}$, where $\{O_r\}$ is a
complete set of positive operators such that $\sum_r O_r=\id$,
namely, the elements of a positive operator valued measurement
(POVM) in the whole subspace of ${\mathscr H}_A\otimes{\mathscr
H}_B$ where the signal states belong. Recalling the invariance of
the Haar measure, $dg=d(gg')$, we can write
\begin{equation}
\langle\carac1\rangle=\sum_r\int dg \;\carac1(g)
|\langle\Phi(g) \ket{\Psi_r}|^2 , \label{chi1 |...|^2}
\end{equation}
where
\begin{equation}
\ket{\Psi_r}\bra{\Psi_r} \equiv U^\dagger_A(g_r)\otimes\id_B \;
O_r\;U_A(g_r)\otimes \id_B. \label{Psi Psi}
\end{equation}
This definition implicitly assumes that optimal POVM's can always
be chosen to have rank one elements~\cite{davies}. We claim that
\textit{(a)} the states of the form~(\ref{state Phi}) are optimal
if the positive coefficients $a_j$ are properly chosen.
\textit{(b)} For the optimal POVM one has
\begin{equation}
\ket{\Psi_r}=\sum_j\ket{\Psi_r^j}=\sum_j\ c_{jr}
\ket{\Phi^j};\quad \sum_r c_{jr}^2=d_j^2. \label{Psi_r}
\end{equation}

To prove~\textit{(a)} we borrow from~\cite{ajv}  some results
concerning the estimation of a $\rm SU(2)$ transformation, in
particular, that the optimal state can be chosen to be
$\ket{\Phi}=\sum_j a_j\ket{\Omega^j}$,  with
%
%
\begin{equation}
\ket{\Omega^{j}}={1\over\sqrt{d_j n_{j}}}\sum_{
m=-j}^j\sum_{\al=1}^{n_{j}}\ket{jm;\al}_A\ket{jm;\al}_B
\end{equation}
instead of~(\ref{state Phi}).
Here $\al$ labels the different $n_{j}$ occurrences of 
$\bf j$ in the Clebsch-Gordan decomposition of $({\bf
1/2})^{\otimes N}$.
We next show that, as far as the evaluation of the maximal
$\langle\carac1\rangle$ (minimal error) is concerned, we need to
consider each $\bf j$  only once. Let us define $v^{(s)}_{\al}$,
$s=1,\dots, n_{j}-1$ as the set of $n_j(n_j-1)$ complex numbers
(which we may regard as the components of $n_{j}-1$ orthogonal
unit vectors) such that $\sum_{\al}
v_{\al}^{(r)*}v_{\al}^{(s)}=\de_{rs}$ and
$\sum_{\al}v_{\al}^{(s)}=0$ (i.e., orthogonal to the
$n_j$-dimensional vector $(1,1,\dots,1)$). We note that  the
states $\ket{\Omega_{s,m}}=\sum_{\al} v^{(s)}_{\al}
\ket{jm;\al}\ket{jm;\al}$ satisfy
$\bra{\Omega^{j}}U_A(g)\otimes\id_B \ket{\Omega_{s,m}}=\sum_\al
v_\al^{(s)}\D{j}{m}{m}(g)/\sqrt{d_jn_j}=0$, for all $ g$, $s$
and~$m$, where we have used that
$\D{j}{m}{m'}(g)=\bra{jm;\al}U(g)\ket{jm';\al}$. Hence,
$\ket{\Omega^{j}}$ effectively lives in only one of the
irreducible representations $\bf j$ and it can be chosen as
in~(\ref{state Phi}) without any loss of generality~$\blacksquare$

To prove claim \textit{(b)} we rewrite~(\ref{chi1 |...|^2}) as
%
\begin{equation}
\langle\carac1\rangle=\sum_r\sum_{jl}a_ja_l\int dg\, \carac1(g)
\bra{\Phi^j\tilde\Phi^l}{\mathscr U}(g)
\ket{\Psi_r^j\tilde\Psi_r^l} , \label{chi1 phi tilde phi}
\end{equation}
where
$\ket{\Phi^j\tilde\Phi^l}=\ket{\Phi^j}\otimes\ket{\tilde\Phi^l}$,
the state $\ket{\tilde\Phi^l}$ is obtained by applying to
$\ket{\Phi^l}$ time reversal only in ${\mathscr H}_A$ (analogous
definitions hold for $\ket{\Psi_r^j\tilde\Psi_r^l}$), and
${\mathscr U}=U_A\otimes U_A\otimes\id_B\otimes\id_B$. By Schur's
lemma, (\ref{chi1 phi tilde phi}) is
\begin{equation}
\langle\carac1\rangle={1\over3}\sum_r\sum_{jl} a_ja_l\; \tr_{\bf
1}(\rho_r^j\otimes\tilde\rho_r^l), \label{chi P1}
\end{equation}
where we have defined
$\rho_r^j=\tr_B(\ket{\Psi_r^j}\bra{\Phi^j})$,
$\tilde\rho_r^l=\tr_B(\ket{\tilde\Psi_r^l}\bra{\tilde\Phi^l})$ and
$\tr_B$ ($\tr_{\bf 1}$) stands for the partial trace over
${\mathscr H}_B$ (over the representation $\bf 1$ invariant
subspace, i.e., $\tr_{\bf 1} O=\sum_{m=-1}^1\bra{1m}O\ket{1m}$).
Using the Schwarz inequality we obtain the bound
\begin{equation}
\tr_{\bf 1}(\rho_r^j\otimes\tilde\rho_r^l)\le\sqrt{\tr_{\bf
1}\left(\rho^j_r\rho^{j\dagger}_r\otimes \id_l\right) \tr_{\bf
1}\left(\id_j\otimes\tilde\rho^l_r\tilde\rho^{l\dagger}_r\right)
},
\end{equation}
where $\id_j$  ($\id_l$) is the identity restricted to the
representation $\bf j$ ($\bf l$) subspace. The equality holds if
$\ket{\Psi_r^j}=c_{jr}\ket{\Phi^j}$ since this choice implies
$\rho^j_r=c_{jr} \tr_B(\ket{\Phi^j}\bra{\Phi^j})=c_{jr}\id_{
j}/d_j$.
To obtain $ \sum_r c_{jr}^2=d_j^2$ one just has to trace~(\ref{Psi
Psi}) on each irreducible representation subspace~$\blacksquare$

With this information we can go back to~(\ref{chi1 |...|^2}) and
cast it as
\begin{equation}
\langle\carac1\rangle\le\sum_r\int dg\,\carac1(g)
 \Bigg|\sum_{j}{a_j c_{jr} \over d_j }\carac{j} (g)   \Bigg|^2 ,
\end{equation}
where we have used that $
\bra{\Phi^j}U_A^\dagger\otimes\id_B\ket{\Phi^j}=\carac{j} (g) /d_j
$. To get rid of the coefficients $c_{jr}$, note that
\begin{equation}
\sum_r c_{jr} c_{lr}\le\sqrt{\sum_r c_{jr}^2
} \sqrt{\sum_r c_{lr}^2\phantom{\big|}} = d_j d_l .
\end{equation}
The equality holds iff
\begin{equation}
c_{jr}=d_j \sqrt{c_r}, \label{cjr=dj cr}
\end{equation}
where $\sum_r c_r=1$. Hence
\begin{equation}
\langle\carac1\rangle\le\int dg\,\carac1(g)
 \Bigg|\sum_{j}{a_j }\carac{j} (g)   \Bigg|^2 .
\end{equation}
The group integral can be easily performed by recalling the
Clebsch-Gordan series
$\carac{j}(g)\carac{l}(g)=\sum_{k=|j-l|}^{j+l} \carac{k}(g)$ and
the orthogonality of the characters~\cite{group}, namely, $\int dg
\carac{j}(g)\carac{l}(g)=\delta_{jl}$. The result can be
conveniently written as
\begin{equation}\label{fidelity-general-3}
\langle\carac1\rangle\le1+ \mathsf{a}^t \mathsf{M}  \mathsf{a}.
\label{1+aMa}
\end{equation}
Here $\mathsf{a}^t=(a_{N/2},a_{N/2-1},a_{N/2-2},\dots)$ is the
transpose of $\mathsf a$, and $\mathsf{M}$ is the tridiagonal
matrix
\begin{equation}\label{matrix}
\mathsf{M}=\pmatrix{0&1&
         &        &       \cr
                    1&\ddots&\ddots
    &\phantom{\ddots} \raisebox{2.0ex}[1.5ex][0ex]{\LARGE
0}\hspace{-.5cm}       &       \cr
                         &\ddots&0    &1  &       \cr
                         & \phantom{\ddots}    &1&0&1
\cr \hspace{.5cm} \raisebox{2.0ex}[1.5ex][0ex]{\LARGE
0}\hspace{-.5cm} &&\phantom{\ddots}&1&\zeta} \ ,
\end{equation}
where $\zeta=-1$  ($\zeta=0$) for $N$ even  (odd). One could also
obtain~(\ref{fidelity-general-3}) directly from~(\ref{chi P1}) by
simply noticing that $\tr_{\bf1}(\id_j\otimes\id_l)=3$ if
$j+l\ge1\ge|j-l|$ and it vanishes otherwise. The maximal value of
the quadratic form in~(\ref{1+aMa}) is given by the largest
eigenvalue of $\mathsf{M}$. Its characteristic polynomial is
$P_n(\lambda)=\det (\mathsf{M}+2\lambda\id)$, where $n$ is the
dimension of $\mathsf M$, namely, $n=N/2+1$ ($n=N/2+1/2$) for $N$
even (odd). Note that we have defined the eigenvalues of $\mathsf
M$ as $-2\la_s$, where $\la_s$, $s=1,2\dots,n$, are the zeroes of
$P_n(\la)$. The characteristic polynomials obey the simple
recurrence relation
\begin{equation}
P_n(\la)=2\la P_{n-1}(\la)-P_{n-2}(\la),
\end{equation}
which is that of the Tchebychev polynomials~\cite{abramowitz}, and
the initial conditions $P_0(\la)=1$, $P_1(\la)=2\la+\zeta$. Hence,
the solution is $P_n(\la)=U_n(\la)+\zeta U_{n-1}(\la)$, where
$U_n(\cos\theta)=\sin[(n+1)\theta]/\sin\theta$ are the Tchebychev
polynomials of the second kind. It is now straightforward to
compute the largest eigenvalue of $\mathsf M$. It can be written
as $2\cos[2\pi/(N+3)]$ and, hence,
\begin{equation}
\langle\carac1\rangle_{\rm max}=1+2\cos{2\pi\over N+3} .
\label{chi1_max}
\end{equation}
One can also verify that the corresponding eigenvector is
\begin{equation}
a_j={2\over\sqrt{N+3}}\sin{(2j+1)\pi\over N+3}  . \label{a_j}
\end{equation}

Eq.~(\ref{chi1_max}) gives an upper bound of the actual
$\langle\carac1\rangle_{\rm max}$. We need to show that this bound
is indeed saturated by a covariant measurement. To do this, we
just trace the conditions under which all the (Schwarz)
inequalities used in the proof are saturated.
Substituting in~(\ref{Psi Psi}) the relation
$\ket{\Psi_r}=\sqrt{c_r}\,d_j\ket{\Phi^j}$, which follows
from~(\ref{Psi_r}) and~(\ref{cjr=dj cr}), we get
\begin{equation}
O_r=c_r\; U_A(g_r)\otimes\id_B\, \ket{\Psi}\bra{\Psi}
\;U_A^\dagger(g_r)\otimes \id_B , \label{finite POVM}
\end{equation}
where $\ket{\Psi}=\sum_{j,m}\sqrt{d_j}\ket{j,m}_A\ket{j,m}_B$. But
for a rescaling factor $c_r$, we see that the positive operators
$O_r$ are all obtained by rotating a fix reference state
$\ket{\Psi}$. This exhibits the covariance of the scheme. An
immediate choice that saturates the bound~(\ref{chi1_max}) is
provided by the continuous POVM
\begin{equation}
O(g)=U_A(g)\otimes\id_B\, \ket{\Psi}\bra{\Psi}
\;U_A^\dagger(g)\otimes \id_B . \label{continuous POVM}
\end{equation}
Using Schur's lemma, we get $\int dg\, O(g)=\sum_j
\id_A^j\id_B^j$, where $\id^j_A$ ($\id^j_B$) is the identity in
Alice's (Bob's)  representation $\bf j$ subspace.
This is the identity in the Hilbert subspace to which all signal
states $\ket{\Phi(g)}$ belong. Hence, the infinite set $\{O(g)\}$
is a POVM for these signal states.

A continuous POVM, such as~(\ref{continuous POVM}), with
infinitely many outcomes is not physically realizable. Hence, it
is important to show that optimal POVMs with a finite number of
outcomes do exist.
 The most
straightforward way of obtaining a finite (though not necessarily
minimal) POVM is by finding a finite set $\{g_r\}$, $r=1,\cdots
n(J)$, of elements of $\mathrm{SU(2)}$ and positive weights
$\{c'_{r}\}$ such that the orthogonality relation
\begin{equation}
\sum_{r=1}^{n(J)}  c'_{r}\,\D{j}{m}{m'} (g_{r}) \Dc{l}{n}{n'}
(g_{r})
 =C_J {\delta^{jl}\delta_{mn}\delta_{m'n'}\over 2j+1}
 \label{discrete-1}
\end{equation}
holds for all $j, l\leq J=N/2+1$, where $C_J=\sum_{r=1}^{n(J)}
c'_r$. This discrete version of the standard orthogonality
relations of $\rm SU(2)$ is only valid up to a maximal value~$J$.
The larger $J$ is, the larger the $n(J)$ that must be chosen.
There are many solutions to these equations and we refer the
reader to~\cite{reference-bbm} for details. Once $\{g_r\}$ and
$\{c'_{r}\}$ have been computed, we simply define $c_r=c'_r/C_J$
and obtain the desired finite POVM elements by substituting these
values in~(\ref{finite POVM}). Eq.~\ref{discrete-1} ensures that
Schur's lemma will work for the finite set $\{g_r,c_r\}$, thus
obtaining $\sum_r O_r=\sum_j\id_A^j\id_B^j$, as it should be.

Let us conclude by summarizing and commenting our results. We
present a covariant (and, hence, very natural) scheme for
transmitting continuous information efficiently through a quantum
channel (the orientation of Alice's reference frame relative to
Bob's). It requires Alice and Bob to share an entangled state of
the form~(\ref{state Phi}). This state can be prepared with, e.g.,
a number of spins or two hydrogen atoms. We determine the
coefficients
---given in~(\ref{a_j})--- which enable Alice to communicate
with the smallest error. The procedure is as simple as
Alice locking her part of the system to her frame and sending it
to Bob, who performs a generalized covariant measurement on the
whole Hilbert space. The error, defined in~(\ref{error}), is given
by
\begin{equation}
\langle h\rangle_{\rm min}=4\left(1-\cos{2\pi\over N+3}\right),
\end{equation}
which follows from the relation $\langle
h\rangle=6-2\langle\carac1\rangle$. The corresponding asymptotic
behavior is $\langle h\rangle_{\rm min}=8\pi^2/N^2$. This is an
striking improvement over any other previously known scheme. We
also prove that the optimal measurements are covariant POVM's,
which one can choose to be either continuous, Eq.~(\ref{continuous
POVM}), or to have a finite number of outcomes.

Our work bears a strong connection with~\cite{ajv}, where the
estimation of  a unitary transformation on qubits is studied. This
problem and that of aligning reference frames are formally the
same.
To be more concrete, let us assume Alice is given a black-box that
performs an unknown unitary operation on qubits (they not need to
be spins in this case) and she is asked to identify it. If she is
allowed to apply the unknown operation $N$ times, the best she can
do is the following~\cite{ajv}: \textit{(a)}~prepare the
$2N$-qubit state~(\ref{state Phi}) \textit{(b)}~apply $u(g)\in
{\bf 1/2}$ over $N$ qubits which results in the
state~(\ref{Phi(g)}) \textit{(c)}~perform the POVM whose elements
are given in~(\ref{continuous POVM}). Note that now all the states
are referred to a unique reference frame: that of Alice (Bob does
not play any role in this case). We must stress that this task
cannot be perform unless both $\ket{\Phi}$ and the POVM elements
can be referred to the same reference frame, which requires that
the person who performs the measurement, if not Alice herself,
must share a reference frame with  her.

Another (minor) difference with respect to the alignment of frames
concerns the figure of merit used in~\cite{ajv}, which is the
fidelity
$F=|\tr[u(g)u^\dagger(g_r)]|^2/4=\carac{1/2}^2(gg_r^{-1})/4$. Our
results can be straightforwardly applied in this context because
of the simple relation $\carac{1/2}^2(g)=1+\carac{1}(g)$. Hence,
for instance, (\ref{chi1_max}) implies that the optimal mean
fidelity is
\begin{equation}
\bar F=\langle F\rangle={1\over2}\left(1+\cos{2\pi\over
N+3}\right),
\end{equation}
whereas for large $N$ one has $\bar F=1-\pi^2/N^2+\dots$. This
extends the results of~\cite{ajv} to arbitrary $N$.

Finally, we would like to point out that our approach resembles
the so called continuous dense coding introduced in
\cite{braunstein}, where the communication of a single phase
---$\mathrm{U(1)}$ group--- was discussed. They found
that dense coding can improve the channel capacity, but not always
does. This is an indication that the absolute optimal scheme for a
phase~\cite{others} does not require bipartite entanglement,
contrasting with our approach for $\mathrm{SU(2)}$, which always
improves the efficiency of the communication.


We are grateful to A. Ac{\'\i}n and E. Jan{\'e} for helpful
conversations. We acknowledge financial support from Spa\-nish
Ministry of Science and Technology project BFM2002-02588, CIRIT
project SGR-00185, and QUPRODIS working group EEC contract
IST-2001-38877.


\begin{thebibliography}{99}
\bibitem{teleportation} C.H.~Bennett {\em et al.}, Phys.~Rev.~Lett.~{\bf 70},
              1895 (1993).
\bibitem{dense} C.H.~Bennett and S.J.~Wiesner, Phys.~Rev.~Lett.~{\bf 69},
              2881 (1992).
\bibitem{direction-1}S.~Massar and S.~Popescu, Phys.~Rev.~Lett.~{\bf 74}, 1259 (1995);
                    R.~Derka, V.~Buzek and A.K.~Ekert, Phys. Rev. Lett.~{\bf 80}, 1571 (1998);
                    J.I.~Latorre, P.~Pascual and  R.~Tarrach, Phys. Rev. Lett.~{\bf 81}, 1351
                    (1998); N.~Gisin and S.~Popescu,
                    Phys.~Rev.~Lett.~{\bf 83}, 432 (1999);
                    S.~Massar, Phys.~Rev.~ A \textbf{62},  040101
                    (2000); A.~Peres and P.F.~Scudo, Phys.~Rev.~Lett.~{\bf 86},
                    4160 (2001); E.~Bagan, M.~Baig and
                    R.~Munoz-Tapia, Phys.~Rev.~A {\bf64},
                    022305 (2001).
\bibitem{direction-bbm} E.~Bagan {\em et al.}, Phys.~Rev.~Lett.~{\bf 85}, 5230
                    (2000); \textit{ibid.}, Phys.~Rev.~A {\bf 63},
                    052309 (2001).
\bibitem{direction-2} D.G. Fischer, S.H. Kienle and M. Freyberger, Phys.
            Rev.~A {\bf 61}, 032306 (2000); R.~D.~Gill and S.~Massar, Phys.~Rev.~A {\bf 61}, 042312 (2000);
            Th.~Hannemann {\it et al.}, Phys. Rev.~A {\bf 65}, 050303 (2002);
             E.~Bagan, M.~Baig and R.~Munoz-Tapia, Phys. Rev. Lett.~{\bf 89}, 277904
             (2002).
\bibitem{reference-ps}A.~Peres and P.F.~Scudo, Phys.~Rev.~Lett.~{\bf 87},
              167901 (2001); N.~H.~Lindner, A.~Peres and D.~R.~Terno,
              Phys.~Rev.~A {\bf 68}, 042308 (2003).
\bibitem{reference-bbm} E.~Bagan, M.~Baig and R.~Munoz-Tapia, Phys. Rev. Lett.~{\bf 87},
          257903 (2001).
\bibitem{ps2}A.~Peres and P.F.~Scudo, J.~Modern Optics \textbf{49}, 1235 (2002).


\bibitem{ariano} G.M.~D'Ariano, P.~LoPresti and M.G.A.~Paris, Phys. Rev. Lett.~{\bf 87}, 270404 (2001).

\bibitem{others} H.~M.~Wiseman and R.~B.~~Killip, Phys.~Rev.~A {\bf 56}, 944
(1997); \textit{ibid}. \textbf{57}, 2169 (1998); D.~W.~Berry,
H.~M.~Wiseman and J.~K.~Breslin, Phys.~Rev.~A {\bf 63}, 053804
(2001);  D.~W.~Berry and  H.~M.~Wiseman, Phys.~Rev.~A {\bf 65},
043803 (2002).

\bibitem{experiments} H.~M.~Wiseman, Phys.~Rev.~Lett.~{\bf 75},
          4587 (1995); W.~P.~Smith \textit{et al}., Phys.~Rev.~Lett.~{\bf 89},
          133601 (2002); M.~A.~Armen \textit{et al}., Phys.~Rev.~Lett.~{\bf 89},
          133602 (2002).


\bibitem{bruss} D.~Bruss and C.~Macchiavello,
Phys.~Lett.~\textbf{A253}, 249 (1999).

\bibitem{bartlett} S.~D.~Bartlett, T.~Rudolph and R.~W.~Spekkens, Phys.~Rev.~Lett.~{\bf 91},
           027901 (2003).

\bibitem{ajv}A.~Acin, E.~Jane and G.~Vidal, Phys. Rev.~A {\bf 64},
             050302 (2001); E.~Jane, Ph.D.~Thesis, University of
             Barcelona, \textit{unpublished}.

\bibitem{holevo} A. S.~Holevo, {\it Probabilistic and Statistical
             Aspects of Quantum Theory} ( North Holland, Amsterdam 1982).

\bibitem{davies} E.~B.~Davies, IEEE Trans.~Inf.~Theory
\textbf{24}, 596 (1978).
\bibitem{group} J. F. Cornwell,  \textit{Group Theory in
            Physics} (Academic Press, London 1984).

\bibitem{abramowitz} M.~Abramowitz and I.~S.~Stegun, \textit{Handbook of
Mathematical Functions} (Dover, New York 1972).

\bibitem{braunstein} S.~L.~Braunstein and H.~J.~Kimble, Phys. Rev.~A {\bf
61}, 042302 (2000).

\end{thebibliography}
\end{document}